# Optical synthesis by spectral translation

Jennifer A. Black[1*], Zachary L. Newman[1,2,3], Su-Peng Yu[1,2], David R. Carlson[1,2,3], and Scott B. Papp[1,2]

**Optical-frequency synthesizers are lasers stabilized and programmed from a microwave clock for applications, especially in fundamental measurements and spectroscopy, optical-communication links, and precision sensing of numerous physical effects. In a synthesizer, a frequency comb provides a reference grid across a broad spectrum and a frequency-tunable laser accomplishes synthesis by phase-lock to the comb. Optical synthesizers have matured to chip-scale with integrated photonics, however, the tunable laser and frequency comb fundamentally constrain the tuning range. Here, we present an optical synthesizer based on four-wave mixing (FWM) spectral translation of a tunable laser and a frequency comb. We implement both the spectral translation and the frequency comb by use of advanced microresonators. The intrinsic energy conservation of FWM ensures deterministic optical synthesis, and it allows a nearly arbitrary frequency tuning range by the dependence of resonant FWM on group-velocity dispersion, temperature, and tunable laser frequency. Moreover, we take advantage of highly efficient parametric amplification associated with spectral translation. We operate spectral translation across output ranges up to 200 THz, and we characterize it through ultraprecise optical-frequency metrology, demonstrating $< 0.1$ Hz absolute accuracy and a fractional frequency precision of $4.7 \times 10^{-13}$ in 1 s measurements. Our experiments introduce integrated-nonlinear-photonic circuits that enable complex intrinsic functionalities, such as our approach to optical-frequency synthesis with nearly limitless bandwidth.**

Optical-frequency combs offer remarkably assured accuracy and precision for frequency metrology, and researchers have been striving to extend their operational bandwidth to as many spectral ranges as possible with sufficient output power for applications. A tunable laser and a frequency comb are the necessary ingredients for a continuous optical-frequency synthesizer. Indeed, such a device readily offers comparable or superior performance to ubiquitous microwave synthesizers, highlighting the potential to leverage the optical domain amongst many modern technologies[1,2]. The modes of an optical-frequency comb are given by $\nu_n = n \cdot f_{rep} + f_{CEO}$, where n is an integer, and $f_{rep}$ and $f_{CEO}$ are microwave frequencies named the repetition frequency and carrier-envelope-offset frequency, respectively. Phase stabilizing $f_{rep}$ and $f_{CEO}$ with respect to a microwave reference clock transfers the stability of the clock to the optical domain. Typically, optical-frequency combs are pumped in the near-infrared, and nonlinear spectral broadening by group-velocity-dispersion (GVD) engineering in bulk and waveguide-based optical systems[3] increases the optical bandwidth. To reach auxiliary spectral windows including XUV, UV and further into the mid infrared, $\chi^{(2)}$ nonlinear processes have been employed including difference-frequency generation and optical-parametric oscillation[4,5]. Demonstrations of continuously tunable optical-frequency synthesizers have employed a fixed-frequency[6] or tunable-frequency[7,8] laser referenced to an optical frequency comb, enabling high optical power but limiting the operational bandwidth.

[1]Time and Frequency Division, National Institute of Standards and Technology, Boulder, CO, USA, [2]Department of Physics, University of Colorado, Boulder, CO, USA., [3]Octave Photonics, Boulder, CO, USA.,*e-mail: jennifer.black@nist.gov

Four-wave mixing (FWM), especially when enhanced by the high-quality factor, small mode volume, and GVD engineering capability of $\chi^{(3)}$ waveguide microresonators, introduces unique opportunities for chip-scale and even fully integrated optical synthesizers[9–13]. Indeed, FWM already offers a proven mechanism to generate soliton microresonator frequency combs, or simply microcombs[14], which are formed from a single continuous-wave (CW) laser with frequency $\nu_0$. The modes of a microcomb are $\nu_\mu = \nu_0 + \mu \cdot f_{rep}$ where $\mu$ is the mode order relative to $\nu_0$. Octave-spanning microcombs [15–17] based on precise GVD engineering in waveguides have been developed, and they have enabled microcomb optical synthesis[18,19] and clock[20,21] demonstrations. Besides microcomb generation that involves many phase-locked modes, optical-parametric oscillation (OPO) and FWM, which involve only three cavity modes, are useful for spectral translation of a laser from one frequency to another. The spectrum of an OPO consists of pump, signal, and idler frequencies, according to $2\nu_p = \nu_s + \nu_i$, and we represent spectral translation via FWM by $\nu_s = 2\nu_p - \nu_i$ in which independent lasers provide the pump and idler. Often signal and idler frequencies are denoted as the high and low frequency components in the FWM process relative to the pump frequency. However, we emphasize that FWM enables spectral translation from either idler to signal frequencies or the reverse. With advancing GVD engineering techniques, the bandwidth of spectral translation, defined by the signal and idler frequency difference, exceeds one octave to cover at least the near infrared, and the fundamental efficiency reaches >10% consistent with OPO in microresonators[22–28]. Moreover, the intrinsic requirement for momentum and energy conservation in OPO and FWM processes favors quantum-limited performance, however, there is no intrinsic stabilization constraint of the spectral-translation bandwidth that provides the essential aspect of a synthesizer. Hence, pumping a microresonator designed for FWM spectral translation with a two-frequency source derived from a microcomb explores a new regime of nonlinear optical interactions.

Here we conceptualize and demonstrate optical synthesis by spectral translation, using a photonic-integrated circuit of microresonators. Our synthesizer uses one microresonator to generate a microcomb and a separate microresonator to implement spectral translation, involving the microcomb mode $\mu_T$ with frequency $\nu_{\mu_T}$ and a pump laser of frequency $\nu_p$ that is offset-frequency-locked to the microcomb. Microresonator GVD engineering provides frequency and phase matching of the microcomb and spectral translation nonlinear processes for efficient conversion. By phase stabilization of the microcomb to a microwave clock and the pump laser with respect to the microcomb, the spectral translation output frequency of our synthesizer $\nu_s = 2\nu_p - \nu_{\mu_T}$ is precisely derived from the clock. We explore the synthesizer's capability for broadband and user-defined tuning, especially the frequency step size that we vary over an extraordinary range from 0.1 Hz to 200 THz to cover vastly different frequency bands. Our experiments utilize an auxiliary, self-referenced frequency comb to assess the precision and accuracy of the optical synthesis chain, including the microcomb, spectral translation, and auxiliary-comb stabilization operations. This work highlights the potential for integrated optical-frequency synthesizers and other nonlinear microresonator circuits through the flexibility, efficiency, precision, and accuracy of FWM.

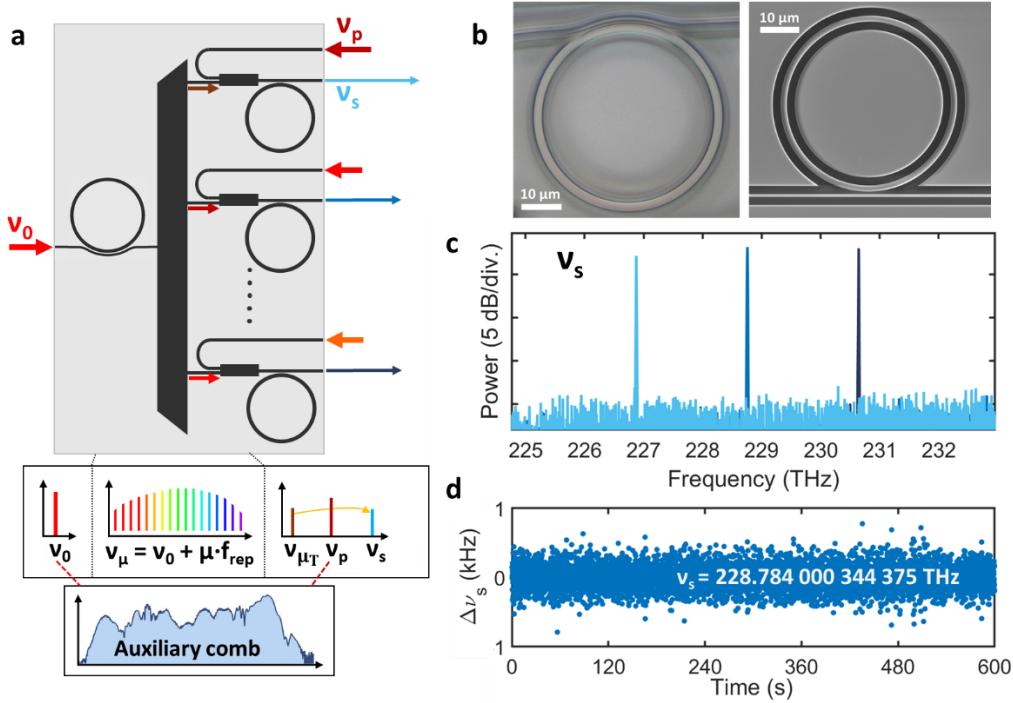

**Fig. 1 | Accurate and precise optical-frequency synthesizer by spectral translation. a,** Concept of a single-chip circuit that implements optical synthesis. A CW pump laser, $\nu_0$, generates a microcomb; modes are demultiplexed and spectrally translated in a second microresonator, using a CW pump laser with frequency, $\nu_p$. **b,** Optical image (left) and electron microscope image (right) of the $Si_3N_4$ and $Ta_2O_5$ microresonators. **c,** Optical-frequency-synthesizer output with tuning of 4 THz (22 nm). **d,** Measured synthesis error $\Delta\nu_s$ (gate time 100 ms).

Optical synthesis by spectral translation centers on the design of FWM phase-matching in microresonators and phase stabilization of the input laser frequencies to generate a target output frequency, $\nu_s$. We implement the optical synthesizer according to the nonlinear microresonator circuit schematic in Fig. 1a, which highlights microcomb wavelength multiplexing to seed different spectral-translation microresonators. Perfect phase-matching for spectral translation requires $\Delta\beta = 2\beta_P - \beta_i - \beta_s = 0$ for the wavevectors of the pump, the idler $\beta_i$ closely linked to the microcomb mode $\mu_T$ that is involved with spectral translation, and the mode associated with the synthesized laser. We derive $\nu_s = \nu_0 + 2(\mu_P \cdot f_{rep} + \delta) - \mu_T \cdot f_{rep}$ for the synthesized output frequency, where $\nu_0$ is the microcomb pump-laser frequency, $f_{rep}$ is the microcomb repetition frequency, and the middle term in parentheses represents the spectral translation pump-laser frequency that is offset locked at a frequency $\delta$ from the microcomb mode $\mu_P$ with respect to the reference clock. A central advantage of our synthesizer is use of pump lasers exclusively in the 1550 nm band; specifically, a CW tunable laser amplified by erbium fiber is the microcomb pump source and a second CW tunable laser pump for spectral translation. Operationally, we design for a choice of $\nu_s$ by adjusting the mode numbers $\mu_P$ and $\mu_T$, and the GVD of the spectral translation microresonator. Moreover, the microcomb $f_{rep}$ is a critical choice; we utilize a 1.002 THz microcomb architecture proven for self-referencing and optical-frequency metrology[17,18]. In any mode of operation, the possibility to sub-divide a large microcomb $f_{rep}$ like 1 THz[18,20] or the use of ultrahigh speed photodetectors and electronics[29] is an important system-design tradeoff. Indeed, in our experiments we electro-optically modulate a portion of the microcomb spectrum in the 1550 nm band for phase stabilization of $f_{rep}$ by feedback to the microcomb pump power. Since FWM conserves energy and hence the phase coherence of the microcomb with respect to the reference clock, we anticipate zero error in the synthesized frequency. However, this is a feature of utmost importance, and we characterize it in detail. Specifically, we use an auxiliary erbium-fiber frequency comb (Appendix) for unambiguous verification by phase-locking $\nu_0$ with respect to the auxiliary comb and measuring an optical-heterodyne beatnote with the synthesized laser output at $\nu_s$.

Our spectral-translation synthesizer (Fig. 1a) can be realized in various microresonator configurations, including with a single chip because microcomb generation and spectral translation are optimized with essentially the same device-layer thickness. In this paper, we demonstrate the flexibility of spectral-translation synthesis by implementing the microcomb with the silicon nitride ($Si_3N_4$ hereafter SiN) platform and spectral translation with the tantalum pentoxide ($Ta_2O_5$ hereafter tantala) platform[30–32]; see Fig. 1b for images of the devices. We create the microcomb with a SiN microresonator that is fully clad in $SiO_2$, 22.67 μm in diameter, resonator waveguide width (RW) of 1653 nm and fabricated by Ligentec on an 770 nm device layer[33]; see Appendix for more details on the microcomb device. We implement custom spectral-translation microresonators with our tantala platform; see Appendix. The microresonators are fabricated with a 570 nm device layer, their radius R of {21, 22, 23} μm corresponds to a free-spectral range of ≈1 THz and varying RW is our principal approach for phase matching. In our experiments, we couple between standard benchtop laser and fiber components and the chips with lensed fiber and inverse-taper waveguides; see Appendix. Since the insertion loss from the microcomb chip to the spectral translation chip is particularly crucial to our experiments, we utilize a thulium fiber amplifier to overcome this loss; see Appendix.

We demonstrate and characterize our spectral-translation synthesizer primarily by its operation of generating a coherent laser output while supporting an exceptionally wide frequency tuning range. Figure 1c demonstrates the optical spectrum of the synthesizer as we step the output laser frequency $\nu_s$ in 2 THz increments through the 1300 nm band. In this experiment, we utilize the same microcomb mode with $\nu_{\mu_T} = 153.868\ 324\ 346\ 2500$ THz, yet we create widely separated outputs by varying the RW of the spectral translation device to change FWM phase matching. Here, the output power of the synthesizer is optimized by minimization of system losses between the chips and thermal tuning of the spectral-translation microresonator, and we generate a sufficient output power for ultraprecise optical frequency metrology. Naturally, the precision we obtain with $\nu_s$ depends exclusively on phase stabilization of the microcomb and $\nu_p$, but conversion efficiency depends on design and fabrication tolerance of the microresonators, which we estimate corresponds to a 10 GHz precision in $\nu_s$. Furthermore, we characterize the absolute optical-frequency-synthesis accuracy by measuring $\nu_s$ with the auxiliary frequency comb. We form an optical heterodyne beat with the synthesizer output and the auxiliary comb and record this frequency with a zero-dead-time frequency counter at 0.1 s gate. Figure 1d presents the absolute synthesizer frequency error $\Delta\nu_s = \nu_b - \nu_s$ where $\nu_b$ is the synthesized optical frequency that we measure with the auxiliary comb. In particular, $\nu_b = f_{CEO} + n_b \cdot f_{rep,aux} + \zeta$, where $f_{CEO}$ and $f_{rep,aux}$ are 30, 250 MHz for the auxiliary comb, $n_b$ is an integer and $\zeta$ is a microwave frequency measured by optical heterodyne of the synthesizer output with the auxiliary comb; see Appendix. Our measurements demonstrate $\Delta \nu_s = +0.035$ Hz in 10 minutes of continuous measurement consistent with zero synthesis error. These results verify sub-Hz accuracy of FWM, phase stabilization of the pump lasers and microcomb, and contributions from the auxiliary comb measurement system.

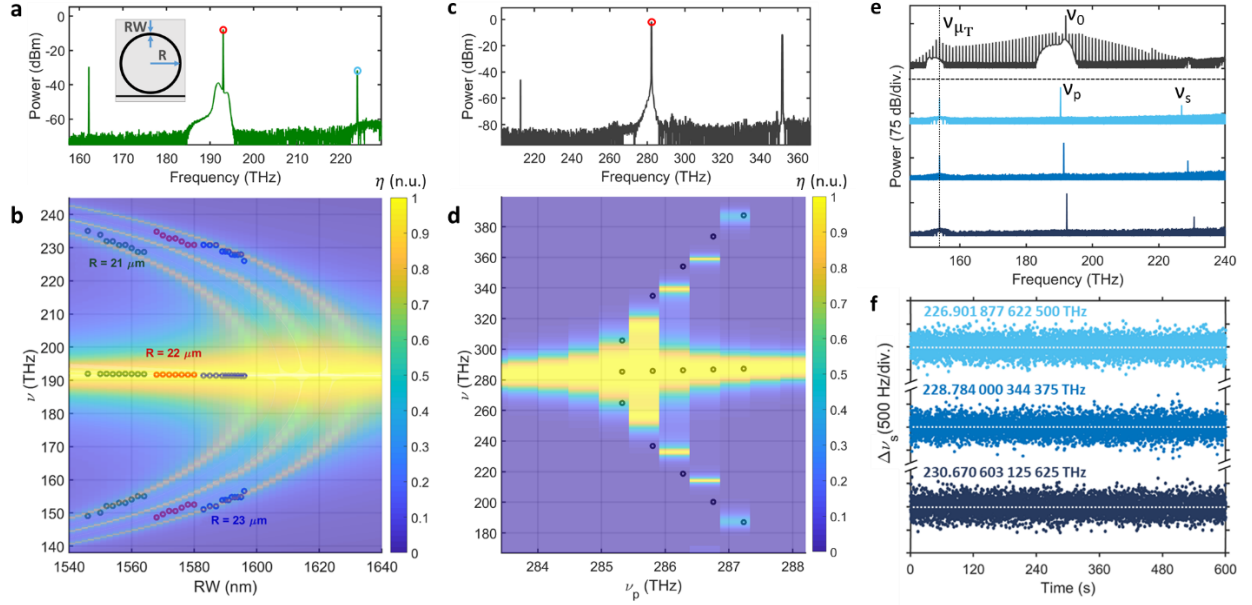

**Fig. 2 | GVD engineering of spectral translation for optical synthesis. a,** Optical spectrum demonstrating efficient spectral translation from 223.7 THz to 162.1 THz. Red and blue circles denote the input pump and signal frequencies. Inset schematic shows the lithographically tunable geometric parameters R and RW. **b,** Green, red, and blue circles denote experimental $\nu_{s,p,i}$ for R = 21, 22 and 23 μm as a function of RW. Heat maps denote the theoretical normalized conversion efficiency ($\eta$). **c,** Optical spectrum demonstrating OPO in a device with R = 45 μm and RW = 1.195 μm. The red circle denotes the input frequency. **d,** Experimental $\nu_{s,p,i}$ for a single microresonator with R = 45 μm and RW = 1.17 μm as a function of $\nu_p$. **e,** Stacked optical spectra of the microcomb and the spectral translation of a single comb tooth ($\nu_{\mu_T}$) to three different frequencies ($\nu_s$) by varying RW. **f,** Measured $\Delta \nu_s$ (gate time 100-ms) of the synthesizer outputs.

Frequency agility, especially by reliable design of FWM and OPO for spectral translation in microresonators, is the key advantage of our synthesizer. Figure 2 explores spectral translation over an optical bandwidth of ≈200 THz and ultraprecise frequency synthesis across the 1300 nm band. Both the microcomb and spectral-translation microresonators operate in the anomalous GVD regime in the frequency range of $\nu_0$ and $\nu_p$, and we design primarily the third- and fourth-order GVD contributions through adjusting the radius (R) and waveguide width (RW) of our resonators; see Fig. 2a and Appendix. In brief, we design the GVD of the microcomb to have zero-crossing wavelengths to generate dispersive waves[34] with one located at $\nu_{\mu_T}$, and we design the GVD of the spectral translation device for energy and momentum conservation at $\nu_{\mu_T}$ and $\nu_s$ with pump frequency $\nu_p$. The integrated GVD describes deviations of a microresonator's free-spectral range ($D_1/2\pi$) at $\nu_p$ with mode order $\mu = 0$[14],

$$D_{int}(\mu) = \omega - (\omega_o + D_1 \cdot \mu) = \frac{D_2\mu^2}{2!} + \frac{D_3\mu^3}{3!} + \frac{D_4\mu^4}{4!} + \cdots.$$

We estimate the dispersive-wave wavelengths by solving the quadratic equation $D_{int}(\mu_{DW}) = 0$, yielding $\mu_{DW} = -\frac{2D_3}{D_4} \pm \sqrt{\left(\frac{2D_3}{D_4}\right)^2 - \frac{12D_2}{D_4}}$. Similarly, energy and momentum conservation enable efficient spectral translation, requiring $D_{int}(\mu_{ST}) = -D_{int}(-\mu_{ST})$ that results in the relationship $\mu_{ST} = \sqrt{\frac{-12D_2}{D_4}}$. Comparing these estimates to results from our experiments is quite instructive as to the frequency determinism of our synthesizer architecture, specifically the dispersive-wave and spectral-translation modes $\mu_{DW} = -38$ and $\mu_{ST} = 40$ are in reasonable agreement with microresonator parameter tolerances for $\nu_0 = 192$ THz and $\nu_p = 191$ THz; see Appendix.

Figures 2a and 2b present spectral-translation characterization with a tunable, 1300 nm probe laser and the $\nu_p$ laser. Figure 2a shows a typical spectral-translation spectrum, and Fig. 2b is a heat map that indicates unit-normalized, spectral-translation efficiency with an overlay of measurement data for discrete devices. To generate an output at any wavelength within our target ≈1270–1330 nm band with a microresonator on a fixed device-layer thickness, we primarily vary RW and use R for fine tuning; see Appendix. In Fig. 2b, the overlay data explores the difference-frequency dependence of $\nu_s$ and $\nu_i$ as a function of RW with a constant $\nu_p$, and we demonstrate the sensitivity of GVD engineering to R. By design of RW, we precisely center $\nu_i$ in the target range and $c/\nu_i$ falls around 1950 nm, and in the most general design case varying R provides an independent control for continuous tuning. We test this by studying three settings of $R = \{21, 22, 23\}$ μm shown by the green, red, and blue circles, respectively, which deterministically shift the phase matching for spectral translation.

Adjusting the pump-laser frequency opens up the widest range of wavelength access by design of microresonator GVD. Figures 2c and 2d present OPO with tantala microresonators and a laser with $\nu_p \approx 280$ THz, demonstrating a span of $\nu_s$ and $\nu_i$ more than 200 THz. Here, we design tantala microresonators of 500 GHz free-spectral range (FSR) and RW for slightly normal GVD in the $\nu_p$ band, eliminating OPO in that frequency range[23]. Figure 2c shows a typical OPO spectrum, and we observe sidebands commensurate with $\nu_s \approx 352$ THz. In Fig. 2d, we present GVD heat maps for OPO in this design regime particularly suited for wideband wavelength access. We step $\nu_p$ through the microresonator modes in the 1064 nm band, realizing variation in the spacing of $\nu_s$ and $\nu_i$. In the future, with finer resolution in fabricated tantala devices features, we could explore a closer connection between GVD and the OPO output.

Bringing together the microcomb, which provides one and the same mode $\mu_T$, and three discrete tantala devices of varying RW for spectral translation, we explore deterministic spectral-translation synthesis across the 1300 nm band (Fig. 2e and f) and highlight the concept of Fig. 1a. Here, the three settings of RW enable 2 THz frequency steps of $\nu_s$, and we utilize slight tuning of $\nu_p$ and microresonator-mode thermal tuning to optimize conversion efficiency. In particular, thermal tuning of the microresonators affects a frequency shift of $-2.3$ GHz/°C and $-0.75$ GHz/°C with the SiN and tantala modes at ≈153.9 THz, respectively. Therefore, we readily achieve optical frequency matching of the microcomb mode and the spectral translation idler mode, but none of these factors that control conversion efficiency impact the frequency stability of $\nu_s$. Indeed, in Fig.

2f, we demonstrate zero $\nu_s$ frequency error determined from 10 min duration, 0.1 s gate-time measurements with the auxiliary comb that are consistent with our understanding of the accuracy and precision of the spectral-translation synthesizer.

The output power of our optical synthesizer by spectral translation is an important metric for applications with synthesized lasers. Overall, the output power depends on the available power of the microcomb mode $\mu_T$ and the conversion efficiency of spectral translation $\eta$. Since energy is conserved in spectral translation, we naturally also expect to observe power gain ($G_{\mu_T}$) of the mode $\mu_T$. An additional parameter of spectral translation is the pump laser power $P_P$. We characterize these aspects of our experiments by calibrating and monitoring the on-chip spectrum, especially at the frequencies $\nu_s$, $\nu_P$, and $\nu_{\mu_T}$; Fig. 3 presents an understanding of $\eta$ and $G_{\mu_T}$ as a function of $P_P$. In our experimental system that couples the microcomb chip, the spectral-translation chip, and the pump lasers, the insertion loss per chip facet is $(\alpha_{\mu_T}, \alpha_P, \alpha_s) = (7.7, 4.7, 5.0) \pm 0.7$ dB. In Fig. 3a, we present the optical spectrum of spectral translation as we vary $P_P$ to characterize conversion efficiency. This data highlights the sparse spectrum of this nonlinear conversion process, involving only the designed waves. Since spectral translation involves two coherent lasers, the pump, and the input from the microcomb, there is no absolute threshold to observe a signal at $\nu_s$. Rather monitoring the entire spectrum provides more information about auxiliary OPO processes, which can reach threshold with sufficient $P_P$. We analyze the spectrum in Fig. 3a to extract $\eta$ and $G_{\mu_T}$, see Figs. 3b and 3c. In particular, we observe relatively high $\eta$ (Fig. 3b) even with $P_P = 7$ mW, and we achieve a maximum conversion efficiency of (-2.6 +/ 1.4) dB with 15 mW. Increasing to $P_P = 19$ mW initiates OPO in microresonator modes near the pump, owing to anomalous GVD, that slightly depletes $\eta$. Commensurate with increasing $\eta$, we also observe gain (Fig. 3b) that peaks at $G_{\mu_T} = 5 \pm 1$ dB. Overall, accounting for $\eta$, the chip coupling losses, and the available off-chip power in microcomb mode of 4 µW, we extract 1 µW. This level of power is suitable to measurements and frequency verification of the synthesizer, and future improvements would lead to milliWatt output power levels. Achieving high output power depends critically on either available microcomb power, since we have already demonstrated ≈ 50% conversion efficiency, or on adjusting our architecture to utilize OPO for spectral translation and stabilization of the OPO frequency spacing to the microcomb.

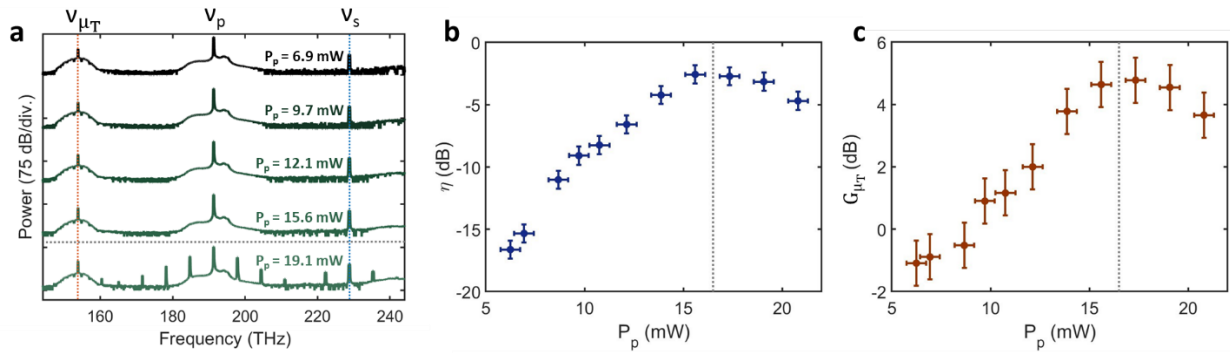

**Fig. 3 | Spectral translation conversion efficiency and gain. a,** Spectral translation optical spectra from $\nu_{\mu_T}$ to $\nu_s$ (red and blue vertical dashed lines) for various on-chip pump powers, $P_p$. The horizontal gray dashed line denotes the threshold for OPO into undesired cavity modes. **b,** $\eta$ as a function of $P_p$ in dB. **c,** Measured gain ($G_{\mu_T}$) on $\nu_{\mu_T}$. The vertical gray dashed line in **b** and **c** corresponds to the horizontal line in **a.**

We expect our optical synthesizer by spectral translation to be phase-coherently linked to the reference clock, which in these experiments is a hydrogen-maser oscillator calibrated with respect to the SI second. Figure 4 explores optical-frequency metrology of the synthesizer, demonstrating its accuracy, precision, and frequency-tuning capability. To assess accuracy and precision, we continuously measure $\Delta\nu_s$ in 0.1 s intervals $\tau(s)$ for a duration greater than two hours by use of a frequency counter, which is referenced to the reference clock; see Fig. 4a. We program the synthesizer to $\nu_s = 228.783\,502\,560\,000$ THz, which we independently verify by characterizing our phase locks of $\nu_0$, $\nu_p$, $f_{rep}$; see Appendix. The Allan deviation (Fig. 4b) of the $\Delta\nu_s$ data demonstrates phase-coherent operation of our synthesizer through its $4.7 \times 10^{-13}/\tau$ fractional-frequency fluctuation dependence over the range $0.1\text{ s} < \tau < 10\text{ s}$. Here, we normalize frequency fluctuations to the ≈75 THz span of spectral translation, and the measurement uncertainty represents sub-Hz accuracy of the synthesizer output for continuous operation greater than two hours. The data record of $\Delta\nu_s$ indicates no frequency error at the level of 83 mHz, ie. 1 part in $10^{16}$ of the synthesizer carrier frequency $\nu_s$, which is less than the 0.1 Hz precision of this experiment. These measurements incorporate the entire optical bandwidth of our spectral-translation synthesizer and self-referenced auxiliary comb, including their electronics for phase stabilization, hence the overall stability of the synthesizer relies on several phase-locks and the optical path length stability of the system. The Allan deviation of Fig. 4b demonstrates phase-stability for over two hours of continuous operation with a deviation from $1/\tau$ behavior for $\tau > 10$ s, which we attribute to fluctuations in the many phase-locks. We note that temperature fluctuations in the room push the locks of $\nu_p$ and $\nu_0$ to unstable ranges of the locking electronics. Also, phase-locking on hours-long time scales leads to drifting of the optical path length and the lensed optical fibers at the chip facets, resulting in signal-to-noise variations due to changing optical power, effecting the stability of the phase-locks.

Our synthesizer supports ultraprecise frequency stepping (Fig. 4c) without any change in phase-stability, GVD, or the temperature of the spectral translation device by adjustment of the offset frequency locks of $\nu_p$ and $\nu_{\mu_T}$. Here, we indicate the frequency programming parameter as $\gamma$, which we apply to either $\nu_p$ and $\nu_{\mu_T}$. With a 3 dB variation in output power, the tuning range is approximately 400 MHz, which is consistent with total linewidth of our resonators. Importantly, $\gamma$ affects an incommensurate magnitude and sign change in $\nu_s$ when applied to either $\nu_p$ and $\nu_{\mu_T}$, according to energy conservation in FWM, $\nu_s = 2\nu_p - \nu_{\mu_T}$. Figure 4c presents three frequency-counter records of $\Delta\nu_s$ in which we program a stepped ramp of $\nu_s$ over a period of several hundred seconds; the three separated panels present different $|\gamma|$ settings 2 kHz, 20 kHz, and 100 kHz, applied at semi-precise 60 s intervals. A value of $\Delta\nu_s = 0$ indicates $\nu_s = 228.805\,087\,882\,250$ THz, and $\Delta\nu_s$ shows the response $2\gamma$ and $\gamma$ that we anticipate as we step $\nu_p$ and $\nu_{\mu_T}$, respectively. We observe no error in $\nu_s$ within the uncertainty of our spectral-translation synthesis chain and auxiliary frequency comb metrology.

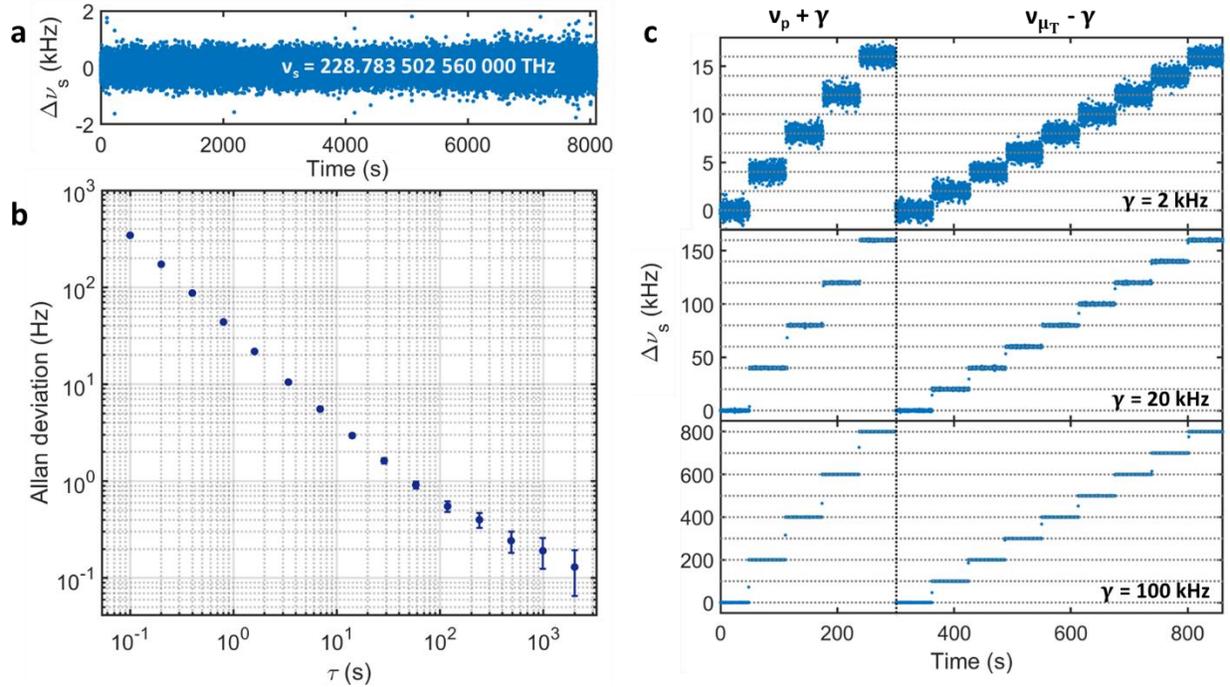

**Fig. 4 | Arbitrary optical frequency synthesis. a,** Measured $\Delta\nu_s$ (gate time 100-ms) for over two hours with a mean $\Delta\nu_s$ of -83 mHz. **b,** Allan deviation of the data in a. At $\tau$ = 1 s, the Allan deviation is 35 Hz indicating fractional stability of the spectral translation across 75 THz of 4.7 x $10^{-13}$. **c,** Measured $\Delta\nu_s$ (gate time 100 ms) as we step $\nu_s$ by changing the lock-point of $\nu_p$ and $\nu_{\mu_T}$ by $|\gamma|$ frequency steps. The gray dashed horizontal lines denote steps of $|\gamma|$ from $\nu_s$.

In summary, we demonstrate an accurate and precise optical-frequency synthesizer by way of spectral translation. This new architecture for optical synthesis establishes nonlinear photonic circuits of microresonators in which frequency-comb generation, spectral translation, and other processes can be combined for system functionality. We leverage the high efficiency and widely controllable FWM phase matching in microresonators; specifically, we demonstrate 50% spectral-translation conversion efficiency of the microcomb and any wavelength access to a range of 200 THz. Our integrated microresonator devices leverage high Q, high nonlinearity, and dimensional control to support designable microcombs and spectral translators. By phase stabilization of the microcomb and the spectral translation pump laser, the synthesized output laser features an Allan deviation of 0.1 Hz, and we explore synthesizer frequency tuning in steps from 0.1 Hz to 2 THz, using microresonator GVD, thermal shifting, and offset locking as actuators without reduced output power. These factors demonstrate a complete set of functionalities for an optical synthesizer with integrated nonlinear microresonators and compatibility with further photonic integration, especially of distributed laser gain. More generally, the concept of an any-wavelength laser enabled by precision control of OPO and FWM illuminates a new paradigm for access to arbitrary portions of the optical spectrum.

## Appendix

### Devices and Nanofabrication

Here we describe the microresonator devices used in our experiments and their fabrication. For the microcomb, we utilize SiN microresonators fabricated by Ligentec according to our designs. A 770 nm thick SiN device layer is deposited on an oxidized silicon wafer by LPCVD, and our design layout for microresonators, waveguide couplers, and other devices is transferred to the SiN layer through DUV stepper lithography and reactive-ion etching. The entire wafer is clad with $SiO_2$ and annealed for several hours at >1000 °C to reduce optical absorption of the SiN and $SiO_2$ materials. A secondary dry etching step separates the wafer to chips. With ≈8 useable DUV stepper fields, ≈30 chips per field, and up to 500 microresonators per chip, we yield a substantial number of devices for experiments on a 4-month duration fabrication cycle. The overall process supports 200 nm feature size, especially for isolated structures like inverse tapers at chip edge. We couple light to the SiN microresonator, using a pulley coupler with pulley length 17 μm, bus waveguide width 500 nm and gap 800 nm that enables efficient broadband coupling. We measure the loaded quality factor at the pump mode near 192 THz to be $\approx 1 \times 10^6$ and we pump the microcomb with ≈55 mW on-chip. For a SiN layer thickness 3% thinner than our target, we find $D_2/2\pi$ = 2.5 MHz, $D_3/2\pi$ = 94.3 kHz and $D_4/2\pi$ = -10.8 kHz yielding $\mu_{DW} = -38, 73$ where the long wavelength dispersive wave is in excellent agreement with experiment. Device dimensions within 5% in thickness and RW are reasonable tolerance for deposition and electron-beam lithography. Sidewall angles can also result in deviations of the GVD from our design geometry.

For the spectral translation microresonators, we utilize tantala microresonators fabricated at NIST[30,35]. A 570 nm thick tantala film is deposited on an oxidized silicon wafer by ion-beam sputtering at FiveNine Optics, and our designs are transferred to the tantala layer through electron-beam lithography and fluorine ICP-RIE. For these devices the upper cladding is air, hence following the etch we separate the wafer into chips with secondary UV laser lithography and dry etching steps. Thermal annealing in air for several hours at 500 °C reduces oxygen vacancies present in the tantala material. Our customized fabrication process yields more than 20 chips with ~100 microresonators per chip in a focused 2-day fabrication period. Given the air cladding, for edge coupling we utilize wide inverse tapers of 2.75 μm. We couple light to the tantala microresonator using straight bus waveguides with width 750 nm and gap 500 nm. We measure loaded quality factors at 193 THz and 224 THz to be $\approx 3 \times 10^5$ and $1 \times 10^6$, respectively. For a tantala microresonator with residual pedestal thickness of 40 nm we find $D_2/2\pi$ = 1.6 MHz, $D_3/2\pi$ = 7.6 kHz and $D_4/2\pi$ = -12.2 kHz yielding $\mu_{ST} = 43$, in good agreement with experiment. For tantala devices, we have found post-processing scanning electron microscope imaging of tantala devices reveal a remaining tantala pedestal with thicknesses ≈10's of nanometers due to incomplete etching through the device layer.

### Benchtop laser systems

A commercially available tunable CW external-cavity diode laser with central wavelength 1550 nm pumps the microresonator for microcomb generation. A single-sideband modulator controls the frequency detuning of the microcomb pump laser, enabling rapid frequency tuning for soliton capture ($\approx$ 6 GHz/100 nm)[36]. After single-sideband modulation and amplification via commercially available erbium-fiber amplifier, the microcomb pump laser goes through a fiber acousto-optic modulator (8 MHz bandwidth) for controlling the optical power before the chip. A second commercially available tunable CW external-cavity diode laser with central wavelength 1550 nm pumps the microresonator for spectral translation and OPO. A commercially available erbium-fiber amplifier amplifies the pump laser, and a wavelength division multiplexer combines the light with a second laser for spectral translation. For characterization of additional spectral translation devices (see Figure 2), we employ two additional widely tunable CW external-cavity diode lasers with central wavelengths at 1050 nm and 1320 nm. We amplify the 1050 nm laser with an ytterbium-doped-fiber amplifier.

We utilize an auxiliary 250 MHz erbium-fiber frequency comb in phase locking the synthesizer and measuring the synthesizer output as described below. The fiber frequency comb is a commercially available mode-locked laser with central wavelength 1560 nm. We amplify and spectrally broaden the frequency comb to octave-spanning for self-referenced detection of $f_{CEO}$ which is phase locked to 30 MHz and we phase lock the fourth harmonic of the directly detectable 250 MHz $f_{rep,aux}$ to a reference frequency synthesizer.

**Design and measurement of spectral translation**

GVD engineering of the microresonators uses finite-element method modelling to predict the wavelength dependent cavity resonances. We fabricate devices with a range of geometric parameter sweeps of R (measured to the center of RW) and RW to cover the full fabrication tolerance range and ensure fabrication of the target GVD microresonator and avoid unwanted higher-order mode crossings. As described in the text, variations in waveguide thickness, residual pedestal, RW and sidewall affect the experimental GVD. We calculate the normalized theoretical spectral translation η of microresonators with various geometries using the cold cavity resonances, ignoring optical power dependent cross- and self-phase modulation of the cavity modes.

Wavelength division multiplexers combine the lasers for launching light onto the chip via lensed optical fibers which produces a spot size of 2 μm at 1550 nm. We precisely align the lensed optical fibers to the chip facets using translation stages and an optical microscope. Experimental measurements of η and $G_{\mu_T}$ were free-running, and we record output optical spectra for various $P_p$ with $\nu_p$ both off- and on-resonance. In these experiments, only $\nu_p$ was frequency tuned and at increasing $P_p$, we slightly cool the tantala ring via a thermoelectric cooler to keep $\nu_{\mu_T}$ on-resonance when $\nu_p$ is on-resonance. Cooling enables counter-acting the thermo-optic coefficient dependent shift of the resonator modes when $\nu_p$ is on-resonance. We measure $G_{\mu_T}$ as the ratio of measured power at $\nu_{\mu_T}$ on the optical spectrum analyzer when $\nu_p$ is on- and off-resonance, respectively. We measure η as the ratio of optical power at $\nu_s$ when $\nu_p$ is on-resonance to optical power at $\nu_{\mu_T}$ when $\nu_p$ is off-resonance. Since we measure higher insertion losses at $\nu_{\mu_T}$ compared to $\nu_s$, we adjust the ratio according to the ratio of relative losses at the two wavelengths.

**Synthesizer phase locking**

We stabilize the synthesizer by phase locking $\nu_0$, $\nu_p$ and $f_{rep}$ to the microwave clock. The pump lasers, $\nu_0$ and $\nu_p$, are phase locked to the self-referenced auxiliary comb described above. Note however that fully self-referenced microcombs have been demonstrated and could be used to implement the synthesizer without the auxiliary comb[19]. We optically heterodyne $\nu_0$ and $\nu_p$ with the auxiliary comb to unambiguously determine their frequency. We use an optical wavemeter to determine the nearest $n^{th}$ comb tooth of the auxiliary comb to $\nu_0$ and $\nu_p$. We measure $f_{rep}$ by passing a portion of microcomb power through an electro-optic comb which produces optical sidebands at a microwave clock referenced synthesizer frequency, $\nu_{EO} \approx 10.018$ GHz[37]. The electro-optic comb generates 100 optical sidebands which fill in the $\approx$1 THz $f_{rep}$. We optically filter a 0.1 nm bandwidth centered between two microcomb modes ($\approx$1541 nm) and measure the optical heterodyne beatnote ($\varepsilon$) of the $50^{th}$ optical sideband from either neighboring microcomb frequency. Then $f_{rep}$ is determined by $100 \cdot \nu_{EO} + \varepsilon$ and we phase lock $f_{rep}$ by controlling the micrcomb pump power with an acousto-optic modulator described above. We phase lock a tunable external-cavity diode laser with central wavelength 1320 nm and milliWatts of optical power to the auxiliary comb to use as a helper laser with frequency $\nu_h$ to boost the signal-to-noise ratio of the final RF beatnote to $\approx$35 dB for frequency counting of $\nu_s$. We measured the in-loop phase locks of $\nu_h$, $\nu_0$, $\nu_p$ and $f_{rep}$ and found all to have frequency error below 600 µHz. The phase lock of $f_{rep}$ had $\Delta\nu_s$ = 93.3 µHz measured over 15 minutes. Multiplying this frequency error across 75 THz of spectral translation yields frequency error < 10 mHz, consistent with our measurements.

**Acknowledgements**

We thank Travis Briles and Andrew Ferdinand for carefully reviewing the manuscript. This research has been funded by NIST, the DARPA DODOS and LUMOS programs as well as AFOSR FA9550-20-1-0004 Project Number 19RT1019.  This work is a contribution of the US Government and is not subject to copyright. Mention of specific companies or trade names is for scientific communication only and does not constitute an endorsement by NIST.